\documentclass[a4paper,11pt]{article}
\usepackage{pos}

\usepackage[sort&compress]{natbib}

\usepackage{hyperref}
\usepackage{multirow}
\usepackage{graphicx}
\usepackage{xfrac}
\usepackage[utf8]{inputenc}

\usepackage{extarrows}

\newcommand{\be}{\begin{equation}}
\newcommand{\ee}{\end{equation}}
\newcommand{\bea}{\begin{eqnarray}}
\newcommand{\eea}{\end{eqnarray}}

\newcommand{\MSb}{\overline{\text{MS}}}

\newcommand{\numax}{\nu_{\textrm{max}}}

\definecolor{Teal}{HTML}{008080}

\let\oldbibliography\thebibliography
\renewcommand{\thebibliography}[1]{\oldbibliography{#1}
\setlength{\baselineskip}{9.5pt}
\setlength{\itemsep}{5pt}} 

\title{Gluon PDF for the proton using the twisted mass formulation of lattice QCD}

\author*[a]{Joseph Delmar}
\emailAdd{jdelmar@temple.edu}
\affiliation[a]{Department of Physics,  Temple University,  Philadelphia,  PA 19122 - 1801,  USA}
\author[b]{Constantia Alexandrou}
\affiliation[b]{Department of Physics, University of Cyprus,  P.O. Box 20537,  1678 Nicosia, Cyprus; Computation-based Science and Technology Research Center,
  The Cyprus Institute, 20 Kavafi Str., Nicosia 2121, Cyprus }
\author[c]{Krzysztof Cichy}
\affiliation[c]{Faculty of Physics, Adam Mickiewicz University, ul.\ Uniwersytetu Pozna\'nskiego 2, 61-614 Pozna\'{n}, Poland}
\author[a]{Martha Constantinou}
\author[b]{Kyriakos Hadjiyiannakou}

\abstract{We present results of the x-dependence of the unpolarized gluon PDF for the proton. We use an $N_f=2+1+1$ ensemble of maximally twisted mass fermions with clover improvement and the Iwasaki improved gluon action. The quark masses are tuned so that the pion mass is 260 MeV. We use a $32^3\times64$ lattice size with a lattice spacing $a=0.093$ fm giving a spatial extent of 3 fm. We employ the pseudo-distribution approach and obtain the light-cone Ioffe time distribution (ITD) combining data for nucleon momentum boosts up to 1.67 GeV and Wilson line lengths, $z$, up to 0.56 fm. We explore systematic effects such as the dependence on the maximum value of $z$ entering the fits to obtain the gluon PDF.}

\FullConference{%
The 39th International Symposium on Lattice Field Theory,\\
8th-13th August, 2022,\\
Rheinische Friedrich-Wilhelms-Universität Bonn, Bonn, Germany
}

\begin{document}
\maketitle
	
\section{Introduction}
\label{sec:intro}
\vspace*{-2mm}

The internal structure of hadrons is an important aspect for understanding the strong force using the theory of quantum chromodynamics (QCD). Due to color confinement, direct observation of quarks and gluons is impossible. Instead, the structure can be separated into a perturbatively-calculable hard-scattering part and a non-perturbative part described by form factors and distribution functions, including parton distribution functions (PDFs). Calculations on a discrete 4-dimensional Euclidean lattice have proven successful in extracting this non-perturbative part, though the light-like nature of these distribution functions makes a direct calculation on a Euclidean lattice impossible. In the last decade, several methods have been proposed to relate lattice data calculated in Euclidean spacetime to physical, light-cone distributions, most notably the quasi-distribution \cite{PhysRevLett.110.262002, Ji:2014sciC} and pseudo-distribution \cite{RADYUSHKIN2017314, PhysRevD.96.034025, RADYUSHKIN2018433, PhysRevD.98.014019, RADYUSHKINintJModPhys} methods, see e.g.\ Refs.~\cite{Cichy:2018mum,Ji:2020ect,Constantinou:2020pek,Cichy:2021lih,Cichy:2021ewm} for recent reviews. Both approaches utilize matrix elements of momentum-boosted hadrons coupled to non-local operators. However, there are significant differences between them. In practice, they are renormalized in differently, with variants of the RI/MOM scheme most often used for quasi-distributions and the ratio scheme for pseudo-distributions. However, the most notable difference is that the factorization into the light-cone counterparts of Euclidean observables is performed either in momentum (quasi) or coordinate space (pseudo). Additionally, reconstruction of the $x$-dependence is typically done employing a fitting ansatz in the pseudo-PDF case. In this work, we use the pseudo-distribution approach to calculate the unpolarized gluon PDF of the proton. Lattice calculations of gluon PDFs present several challenges beyond the need to effectively match the lattice data to the light cone. The gluon component is a purely disconnected diagram, resulting in significant noise and requiring at least an order of magnitude more statistics compared to the quark counterpart. There is also unavoidable mixing with the quark singlet PDFs that must be addressed.

The gluonic component of hadron structure has received less attention than that of the quark. However, gluons contribute significantly to various physical quantities. For instance, phenomenological data and lattice calculations suggest the gluons account for approximately $40\%$ of the hadron's momentum at a scale of $6.25 \, \text{GeV}^2$ \cite{PhysRevD.89.054028, PhysRevD.96.054503}. A better understanding of how the gluon contributes to hadron structure is essential. There are dedicated lattice~\cite{Fan:2018dxu,Fan:2020cpa,HadStruc:2021wmh,HadStruc:2022yaw} and phenomenological analyses of experimental data sets~\cite{NNPDF:2017mvq,Hou:2019efy,Moffat:2021dji} to understand the gluonic structure of the proton. The lattice data on $x$-dependent quantities have the potential to assist by constraining global analyses as done for the case of quark PDFs~\cite{PhysRevD.104.016015,NNPDF,PhysRevD.93.114017}.

In these proceedings, we present our calculation of the unpolarized gluon PDF for the proton using the pseudo-PDF approach. We compare our results with an existing lattice calculation from the HadStruc collaboration \cite{HadStruc:2021wmh} and global analysis from the JAM collaboration \cite{PhysRevD.104.016015}.


\vspace*{-2mm}
\section{Methodology}
\label{sec:setup}
\vspace*{-2mm}

The calculation relies on matrix elements of a non-local gluon operator that couples to proton states, $N(P)$, that are boosted with momentum $P$. The operator related to the gluon PDFs is non-local and constructed by two gluon field-strength tensors, $F^{\mu\nu}$, separated by spatial distance $z$, and two straight Wilson lines, connecting points $0\to z$ and $z\to0$,
\begin{equation}
    \label{eq:gluon_oper}
    M_{\mu i; \nu j}(P,z)  = \langle N(P)| F_{\mu i}(z) W(z, 0) F_{\nu j}(0)  W(0,z)|N(P) \rangle \,,
\end{equation}
where $F_{\mu\nu}$ is the gluon field strength tensor defined as
{\small{
\begin{eqnarray}
    F_{\mu\nu} (x) &=& \frac{i}{8 g_0} \bigg[ U_\mu(x) U_\nu(x+a\hat{\mu}) U^\dag_\mu(x+a \hat{\nu}) U^\dag_\nu(x) + U_\nu(x) U^\dag_\mu(x+a\hat{\nu}-a\hat{\mu}) U_\nu^\dag(x-a\hat{\mu}) U_\mu(x-a\hat{\mu}) \nonumber \\
    && \qquad + U^\dag_\mu(x-\hat{\mu}) U^\dag_\nu(x-a\hat{\nu}-a\hat{\mu}) U_\mu(x-a\hat{\nu}-a\hat{\mu}) U_\nu(x-a\hat{\nu}) \nonumber \\
    && \qquad+ U^\dag_\nu(x-a\hat{\nu}) U_\mu(x-a\hat{\nu}) U_\nu(x-a\hat{\nu}+a\hat{\mu}) U^\dag_\mu(x) - h.c \bigg]\,,
    \label{Eq:FST}
\end{eqnarray}
}}
with $g$ being the bare coupling constant. Note that the two Wilson lines are needed to make the operator gauge invariant. The matrix elements depend on the Lorentz indices $\mu\,,\nu,\, i,\, j$, which can be temporal or spatial. 
The various options of the indices lead to the construction of operators with different properties. Here, we use the operator
\begin{eqnarray}
{\cal O}_3 &\equiv& \frac{1}{2} \sum_{i} F_{it}(x+z \hat{z}) W(x+z \hat{z},x) F_{iz}(x) W(x,x+z \hat{z})\,, \quad i \ne z\,,
\end{eqnarray}
which is free of mixing under renormalization but has a non-vanishing vacuum expectation value that must be subtracted. 
We note that, regardless of the choice discretization, the unpolarized gluon PDF mixes with the unpolarized singlet quark PDF. 
The matrix elements $M_{\mu i; \nu j}$ denote the ground state, which is extracted from the ratio between three-point and two-point correlation functions,
\begin{equation}
 R(t,\tau,t') = \frac{C^{\text{3pt}}(t_s,\tau; {\bf P})}{C^{\text{2pt}}(t_s; {\bf P})} \stackrel{t_s<\tau<t'} \to  \frac{4}{3} \left( \frac{m^2}{4 E} - E \right) M\,,
  \label{Eq:Ratio}
\end{equation}
where we drop the indices of $M$ for simplicity, and $t_s$ and $\tau$ denote the time of the sink and
operator insertion, respectively. Without loss of generality, the source is taken at zero time. $C^{\text{3pt}}$ contains disconnected contributions, which are constructed by the expectation value of a product of a gluon loop  with the proton two-point function. For the unpolarized gluon PDF, the parity projector $\Gamma_0 \equiv \frac{1}{4} (1+\gamma_0)$ is applied to both the three- and two-point functions.

There are a number of nontrivial steps required to extract the $x$-dependence of the gluon PDF. Here, we implement the pseudo-ITD framework, which begins with the construction of appropriate ratios of the matrix elements,
\begin{equation}
    \label{eq:double_ratio}
    \mathfrak{M}(\nu,z^2) \equiv \bigg( \frac{M(\nu,z^2)}{M(\nu,0)|_{z=0}} \bigg) \bigg/ \bigg( \frac{M(0,z^2)|_{p=0}}{M(0,0)|_{p=0,z=0}} \bigg)\,.
\end{equation}
Note that the matrix elements are expressed in terms of the Wilson line length, $z$, and Ioffe time, $\nu=z\cdot P$.
For multiplicatively renormalized operators, the above ratio, the so-called reduced Ioffe-time distribution (pseudo-ITD), cancels the divergences, including the power divergence due to the presence of the Wilson line. The use of the ``double ratio'' has been proven essential in suppressing discretization effects and higher-twist effects, which are assumed similar in the two single ratios shown above~\cite{Orginos:2017kos}. Since $\mathfrak{M}$ serves as a nonperturbative renormalization prescription, it is governed by the scale $1/z$, which is related to the renormalization scale, $\mu^2$.

One must apply a matching procedure to $\mathfrak{M}$ to extract their light-cone counterpart, $\mathcal{Q}$, which is known to one loop,
\begin{equation}
    \label{eq:itd_equation}
    \mathcal{Q}(\nu,z^2,\mu^2) = \mathfrak{M} + \frac{\alpha_s N_c}{2\pi}\int_0^1 du \; \mathfrak{M}(u\nu,z^2) \bigg\{ \text{ln}\bigg(\frac{z^2 \mu^2 \text{e}^{2\gamma_E}}{4}\bigg)B(u) + L(u) \bigg\},\,
\end{equation}
where
\begin{equation}
    \label{eq:B_Lkernel}
    B(u) = \bigg[\frac{1+u^2}{1-u}\bigg]_+\,,\qquad 
    L(u)= 4 \bigg[\frac{u+\text{ln}(\overline{u})}{\overline{u}}\bigg]_{+} + \frac{2}{3} \big[1-u^3\big]_{+}\,,
\end{equation}
and the plus prescription is given by
$\int^1_0 [f(u)]_+ Q(u\nu)=\int^1_0 f(u) (Q(u\nu)-Q(\nu))$.
The matching equations involve evolving the reduced-ITD to a common scale ($B(u)$ term) and converting the expressions to the light-cone ITD in the $\rm \overline{MS}$ scheme ($L(u)$ term). It is convenient to rewrite the inverse of Eq.~\eqref{eq:itd_equation} in two parts so that one can perform the evolution and scheme conversion separately to study their effect in the final ITD,
\begin{eqnarray}
\label{eq:evol}
\mathfrak{M}'(\nu,z^2,\mu^2) = \mathfrak{M}(\nu,z^2) - \frac{\alpha_s C_F}{2\pi} \int_0^1 du 
\times \ln \left(z^2\mu^2 \frac{e^{2\gamma_E +1}}4\right) 
 B(u) \mathfrak{M}(u\nu,z^2)\,,
\end{eqnarray}
where $\mathfrak{M}'(\nu,z^2,\mu^2)$ is the evolved ITD, which depends on $\nu$, the final scale $\mu$ and the initial scale $z$. Finally, conversion to the $\overline{\rm MS}$ scheme is given by
\begin{equation}
\label{eq:match}
Q(\nu,z^2,\mu^2) = \mathfrak{M}'(\nu,z^2,\mu^2) - \frac{\alpha_s C_F}{2\pi} \int_0^1 du  L(u)  \mathfrak{M}(u\nu,z^2)\,.
\end{equation}
$Q$ is averaged over the same values of $\nu$ extracted from different combinations of $P$ and $z$. In preparation for extracting the gluon PDF, the ITD is then fitted according to the minimization of
\begin{equation}
    \label{eq:chi_sq}
    \chi^2 = \sum_{\nu=0}^{\nu_{\rm max}} \frac{\big(Q(\nu,\mu^2)-Q_f(\nu, \mu^2)\big)^2}{\sigma^2_Q(\nu, \mu^2)}\,,
\end{equation}
where $\sigma_Q^2(\nu,\mu^2)$ is the statistical error of the light-cone ITD $Q(\nu,\mu^2)$.
Once the light-cone ITD has been extracted, one may obtain the light-cone PDF, which is related to the ITD via the Fourier transform 
\begin{equation}
\label{eq:PDF2ITD}
Q(\nu,\mu^2) =\int_{0}^1 dx \, \cos(\nu x) x g(x,\mu^2)\,.
\end{equation}
The reconstruction of $g(x,\mu^2)$ poses an inverse problem~\cite{Karpie:2018zaz} because the inverse equations are ill-defined due to the limited number of lattice data for $Q(\nu)$. The main challenge is that the lattice data are obtained on a relatively small number of momenta $P$, and, thus, the range of Ioffe time spans from 0 up to some $\numax$. Therefore, to extract $g(x,\mu^2)$, one requires additional information, which can be chosen in several ways. Here, we reconstruct the gluon PDF by using a fitting ansatz commonly used in the analysis of experimental data sets, that is
\begin{equation}
\label{eq:ansatz}
x q(x) = N x^a (1-x)^b,
\end{equation}
where the exponents $a,\,b$ are fitting parameters and $N$ is the normalization that is fixed by the gluon momentum fraction $\int_0^1 dx \,x g(q) = \langle x \rangle_g$.


\subsection{Lattice Calculation}
\label{sec:latt_details}
\vspace*{-2mm}

The calculation of the gluon PDF performed here is done using an $N_f=2+1+1$ ensemble of twisted-mass clover fermions, and Iwasaki improved gluons generated by the Extended Twisted Mass Collaboration (ETMC)~\cite{Alexandrou:2018egz}. The quark masses of this lattice are such that the pion has approximately twice its physical mass ($m_\pi = 260$ MeV). The lattice spacing is $a=0.0938(2)(3)$ fm and the lattice volume is $32^3 \times 64$. 

The matrix elements are calculated with protons at rest, as well as with four values of the momentum boost, that is, $P=0.42,\, 0.83,\, 1.25,\, 1.67$ GeV. To benefit from the correlations between the numerator and denominator of the reduced ITD, we obtain the matrix element $M$ at the same configurations and with the same source positions for all values of $P$. To increase the statistics by a factor of six, we calculate $M$ with the Wilson line and momentum boost in the $\pm x,\,\pm y,\,\pm z$ directions, which we averaged over as they lead to the same PDF. This is important, as the statistics required for gluonic quantities is much higher than for the quark counterpart, due to the increased gauge noise in the correlator. By construction, the disconnected contributions are evaluated at open sink time, and we investigate excited-states contamination by varying the source-sink time separation.

To extract the gluon PDFs,  one needs to use smoothing techniques, and here we use stout smearing~\cite{Morningstar:2003gk} on the gauge links entering the field strength tensor and the Wilson line independently, with parameter $\rho=0.129$~\cite{Alexandrou:2016ekb,Alexandrou:2020sml}. We apply a 4D smearing at 10 and 20 steps ($N^{\rm F}_{\rm stout}$) for the field strength tensor, while for the Wilson line we apply 3D smearing with 0 and 10 steps ($N^{\rm W}_{\rm stout}$) .
To improve the overlap with the proton ground state, we apply momentum smearing~\cite{Bali:2016lva} at an optimized value of its parameter, $\xi=0.6$, for the three highest momentum boosts, $P=0.83,\, 1.25,\, 1.67$ GeV. The momentum smearing technique has been proven essential in suppressing the gauge noise in matrix elements with boosted hadrons and non-local operators~\cite{Alexandrou:2016jqi}. The nature of the gluon calculation requires significant statistics to reduce errors and provide meaningful results. To this end, we produce 200 source positions for each configuration.
In Table~\ref{tab:stat}, we summarize the statistics for this calculation.

\begin{table}[h!]
\begin{center}
\renewcommand{\arraystretch}{1.4}
\begin{tabular}{c|ccccc}
\hline
$P_3$ [GeV]  &  $\quad N_{\rm confs} \quad$ &  $\quad N_{\rm src} \quad $  &  $\quad N_{\rm dir}\quad  $ & $\quad N_{\rm meas}\quad $\\
\hline
0, 0.42, 0.83, 1.25, 1.67 & 1,134 & 200 & 6 & 1,360,800\\
\hline
\end{tabular}
\vspace*{0.15cm}
\caption{Total statistics for the calculation for each value of $P_3$. $N_{\rm confs}$ is the number of configurations, $N_{\rm src}$ the number of source positions, $N_{\rm dir}$ is the number of spatial directions for the Wilson line and $P_3$, and $N_{\rm meas}$ is the number of total measurements ($N_{\rm meas} = N_{\rm confs} \times N_{\rm src} \times N_{\rm dir}$). }
\label{tab:stat}
\end{center}
\end{table}

\vspace*{-6mm}
\section{Results}
\label{sec:results}
\vspace*{-2mm}

\begin{figure}[h!]
    \centering
    \includegraphics[scale=0.5]{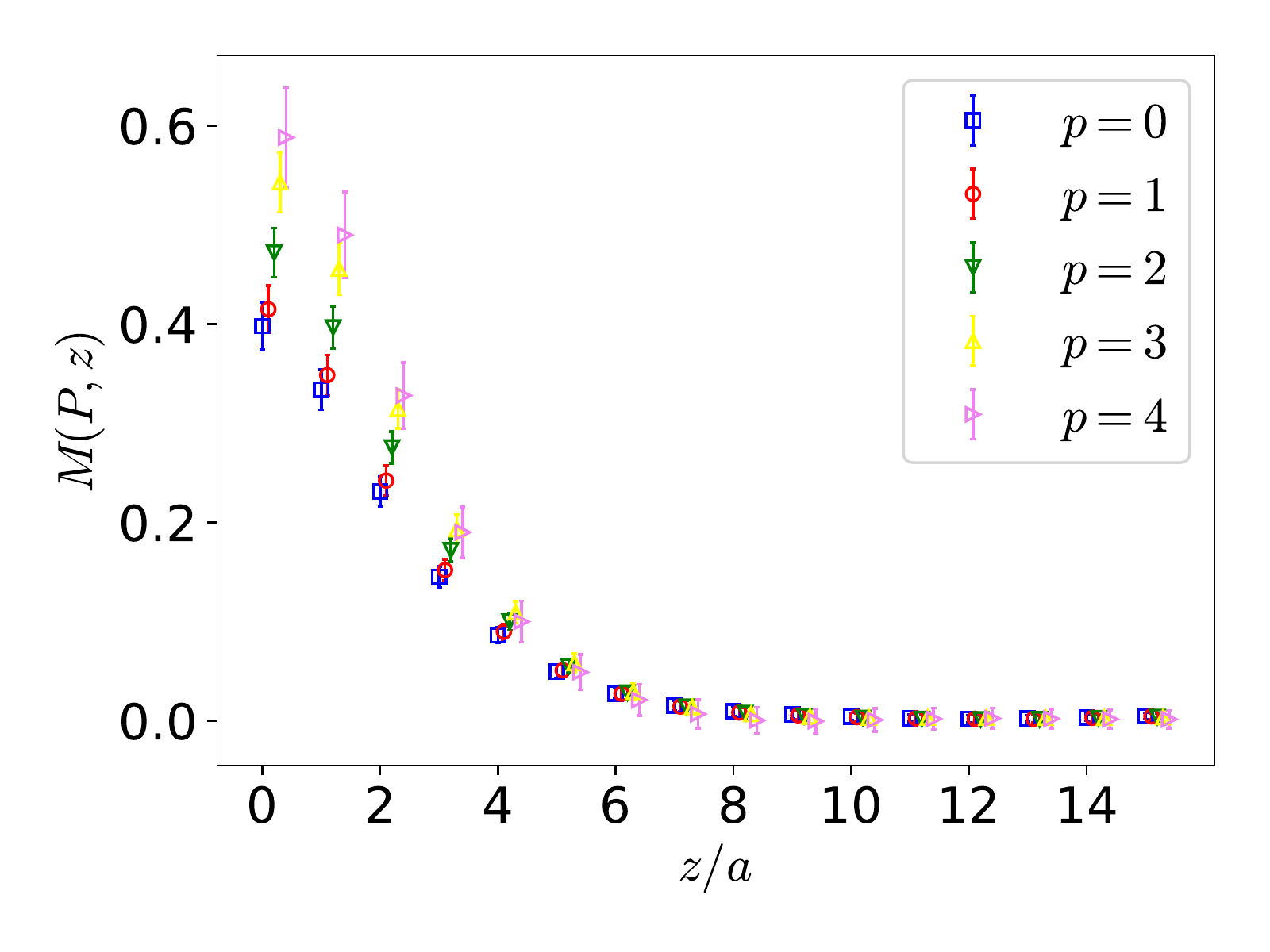}
    \vspace*{-0.5cm}
    \caption{\small{Left: Matrix elements of Eq.~\eqref{eq:gluon_oper} as a function of the length of the Wilson line, $z/a$. The data at momentum boost $P=\frac{2\pi}{L} p$ with $p=0,\,1,\,2,3,4$ are shown with blue squares, red circles, green downward-pointing triangles, yellow upward-pointing triangles, and magenta rightward-pointing triangles, respectively. }}
    \label{fig:gluon_matrix}
\end{figure}

We begin the presentation with the bare matrix elements, $M$, as extracted from Eq.~\eqref{Eq:Ratio}, including all kinematic factors. In Fig.~\ref{fig:gluon_matrix}, we compare the data for all values of the momentum boost using $t_s=9 a$,  $N^{\rm F}_{\rm stout}=20$, and $N^{\rm W}_{\rm stout}=10$. The behavior of the data is as expected with an increase of $P$, that is, the signal quality decreases. We find that the relative error at $z=0$ for $P=0$ is $\sim$6\%, while for $P=1.67$ GeV becomes $\sim$9\%. We remind the reader that the statistics is the same for all momenta.  the maximum value of the matrix element is at $z=0$ and it decays as $z$ increases. We also find that at $z \sim 8a$, the matrix elements decay to zero.

\begin{figure}[h]
\hspace*{-0.5cm}
    \includegraphics[scale=0.4]{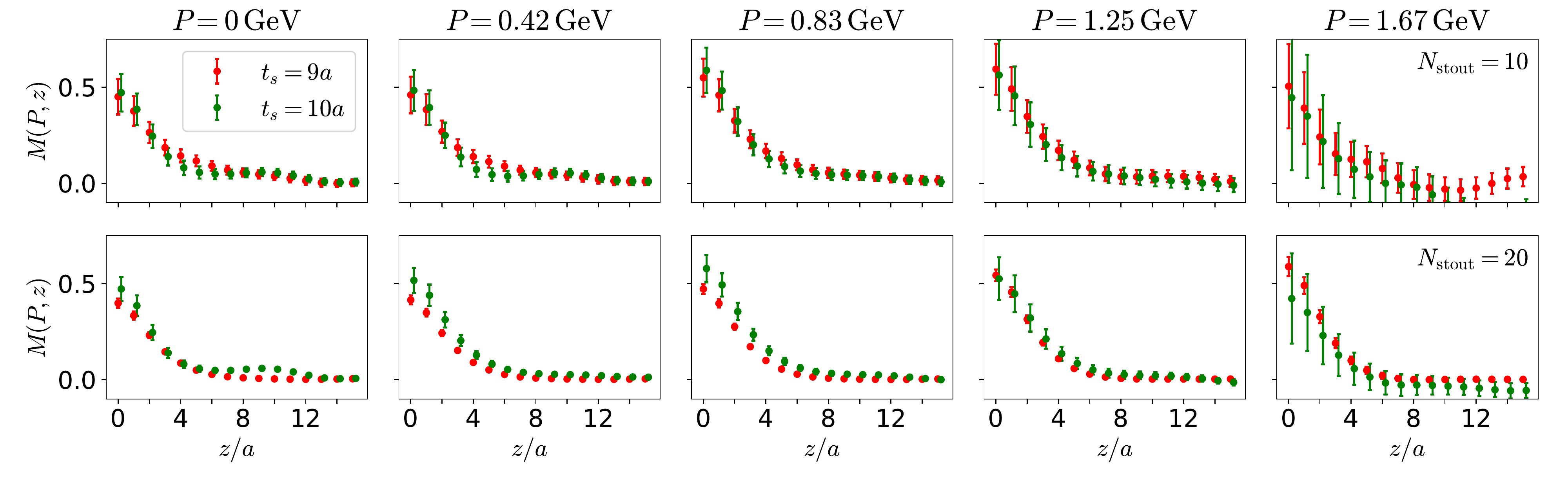}
    \vskip -0.5cm
    \caption{Matrix elements for $N^{\rm F}_{\rm stout}=10$ (top row) and  $N^{\rm F}_{\rm stout}=20$ (bottom row)  at each momentum boost (columns). The values at $t_s=9a$ are shown in red and at $t_s=10a$ in green.}
    \label{fig:ME_stout_comp}
\end{figure}

We find that the introduction of stout smearing improves signal quality. As a representative example, we compare the two values used for the gauge links of the field strength tensor, 10 and 20 steps. The results are shown at the top and bottom row of Fig.~\ref{fig:ME_stout_comp}, respectively. We show data at two values of the source-sink time separation, $t_{s}=9a$ and $t_{s}=10a$. There are several observations from the above-mentioned comparison: (a) $N_{\rm stout}=20$ leads to statistically more accurate results for both $t_{s}=9a$ and $t_{s}=10a$; (b) indication of excited-states effects is found for $N_{\rm stout}=20$, while for $N_{\rm stout}=10$ excited states are hidden in the statistical uncertainties; (c) the statistical noise is enhanced as the momentum boost increases. Based on the above, we process with $N^{\rm F}_{\rm stout}=20$ for the final analysis.

\begin{figure}[h!]
    \centering
    \includegraphics[scale=0.41]{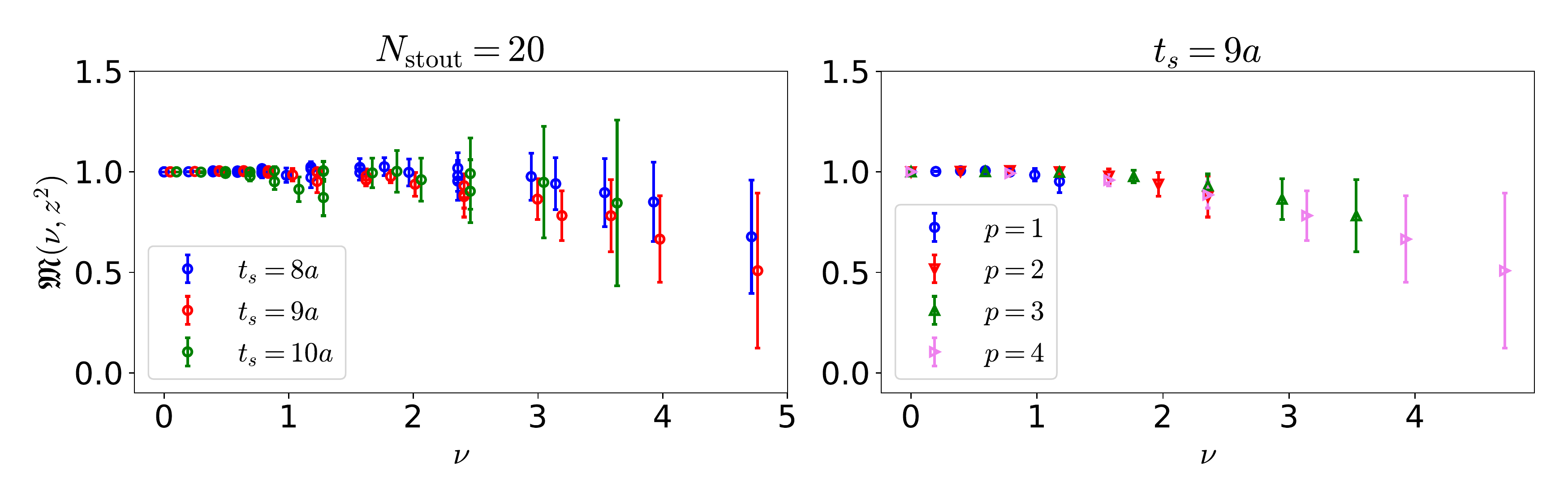}
        \vskip -0.5cm
    \caption{Reduced-matrix elements vs. Ioffe time. Left panel: the values for $t_s=8a$ (blue), $t_s=9a$ (red), and $t_s=10a$ (green) for $N_{stout}=20$ with the highest boost excluded from $t_s=10a$ for clarity. Right panel: final $t_s=9a$ values with boosts $P=\frac{2\pi}{L}p$: $p=1$ (blue circles), $p=2$ (red down-pointing triangles), $p=3$ (green up-pointing triangles), and $p=4$ (violet right-pointing triangles).}
    \label{fig:DR_tsink_test}
\end{figure}

The matrix elements presented in Fig.~\ref{fig:gluon_matrix} are the ingredients entering the double ratio of Eq.~\eqref{eq:double_ratio}. It is interesting to investigate excited-states effects in the reduced ITD, which is the core element for the pseudo-distribution analysis. In Fig.~\ref{fig:DR_tsink_test}, we show $\mathfrak{M}$ for three values of the time separation, $t_s=8a,\,9a,\,10a$. We find that the statistical error increase is sizeable between $t_s=9a$ and $t_s=10a$ and the signal is already lost at $t_s=12a$; the latter is not shown here. Based on these results, we assess that excited states are within the statistical uncertainties, and we choose $t_s=9a$ to proceed with the rest of the analysis, which is shown in the right panel of Fig.~\ref{fig:DR_tsink_test}. We note that only the matrix elements up to $z=6a\sim0.56$ fm are included, which allows us to extract $\mathfrak{M}$ up to about $\nu=5$. The $P$-dependence of $\mathfrak{M}$ is found to be very small, as data at the same $\nu$ but different $z,\,P$ are compatible within errors. As a result, $\mathfrak{M}$ is a smooth function in Ioffe time, which allows for a controlled interpolation needed for the scaling and matching procedure.  

\vspace*{-0.5cm}
\begin{figure}[h!]
    \centering
    \includegraphics[scale=0.48]{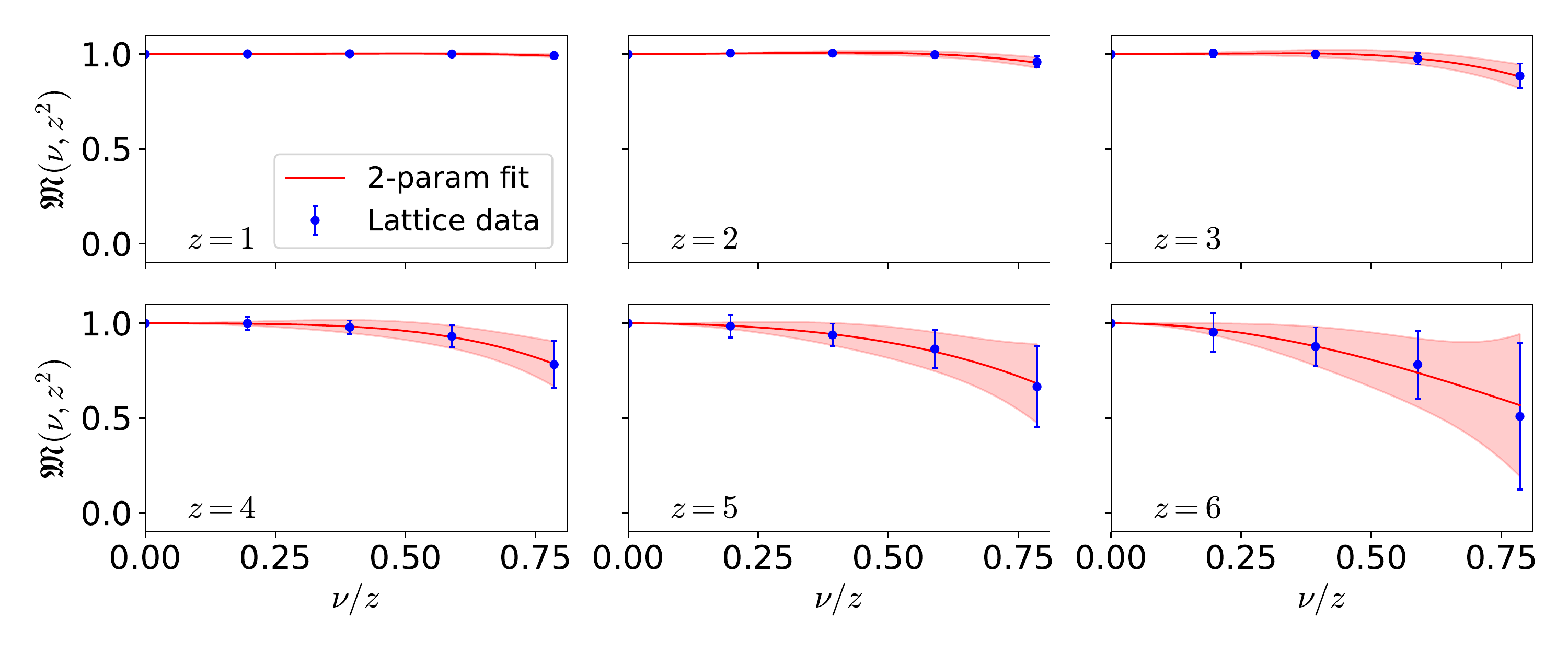}
    \vskip -0.5cm
    \caption{Lattice data of the reduced ITDs for $z=1a - 6a$ (blue points) and their interpolation at fixed $z^2$ using a second order polynomial fit (red bands).}
    \label{fig:DR_interpolation}
\end{figure}

The reduced ITD are interpolated at each value of $z^2$ and varying $\nu$ to obtain a continuous function in $\nu/z$, needed to perform the matching procedure. We tested a linear and a second-order polynomial fit. While the polynomial fit proves to be the best suited for the procedure of matching to the light-cone PDF, the choice is mostly irrelevant at very small values of $z$, as can be seen in Fig.~\ref{fig:DR_interpolation}.

\begin{figure}[h!]
    \centering
    \includegraphics[scale=0.43]{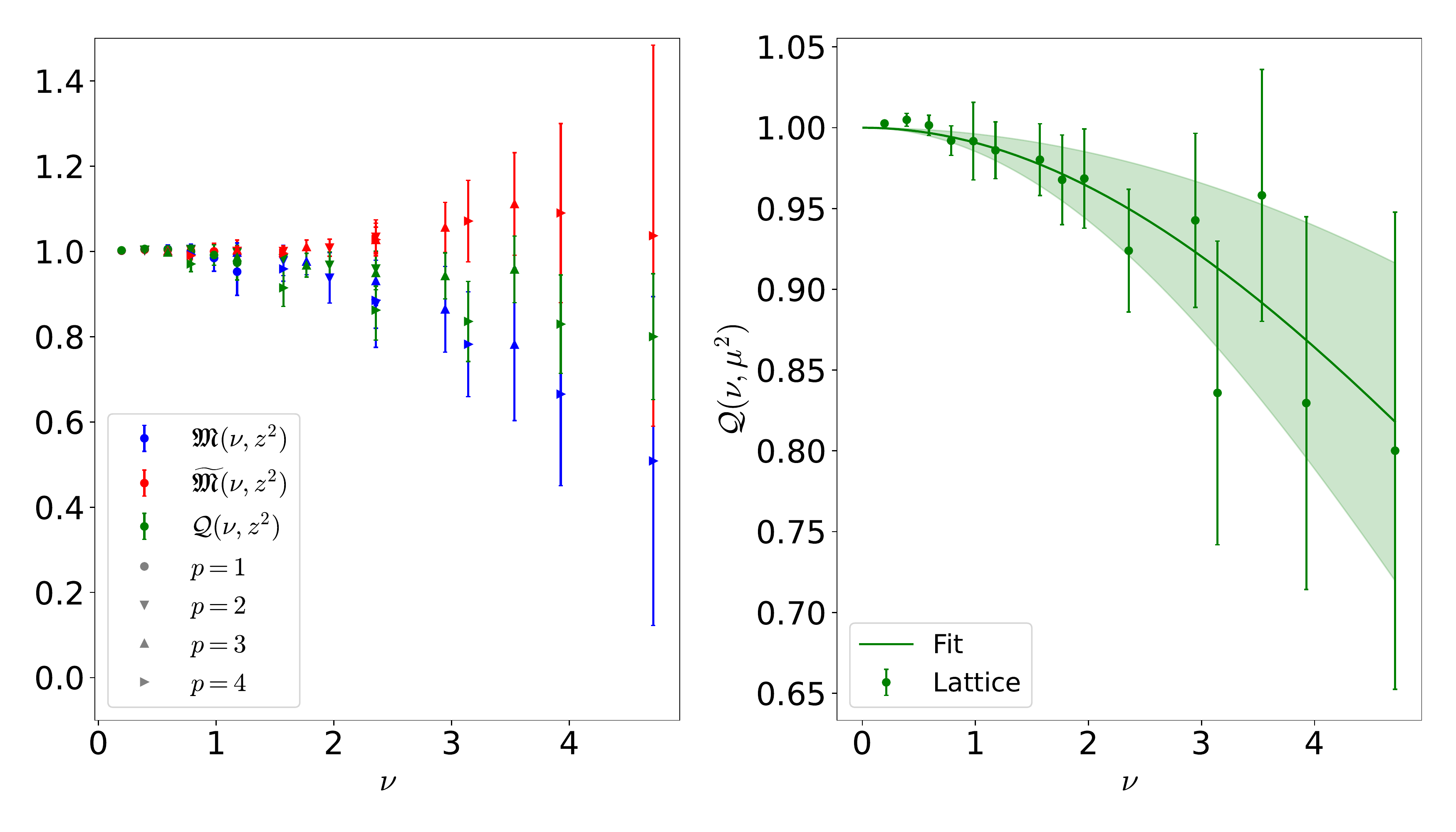}
    \caption{Left: The reduced (blue), evolved (red), and matched (green) ITDs shown for momentum boosts $p=1$ (circles), $p=2$ (down-pointing triangles), $p=3$ (up-pointing triangles), and $p=4$ (right-pointing triangles), where $p=P\frac{L}{2\pi}$. Right: The final light-cone ITD and its fit.}
    \label{fig:evolved_ITD}
\end{figure}

The extraction of the light-cone ITDs can be performed in two steps, as given in Eq.~\eqref{eq:itd_equation}; that is, the evolution to a common scale chosen to be 2 GeV and converting the expressions to the light-cone ITD in the $\rm \overline{MS}$ scheme. The former is done using the evolution kernel $B(u)$, while $L(u)$ is the conversion to the $\rm \overline{MS}$ scheme. Both expressions are  given in Eq.~\eqref{eq:B_Lkernel}.
The resulting evolved- and matched-ITDs are shown in Fig.~\ref{fig:evolved_ITD}. We find that the evolution increases the values of the evolved ITD relative to those of the reduced-ITD, while the matching lowers the values and brings the final light-cone ITD to be compatible with the initial reduced ITDs within error bars. In all cases, the values from different $(P,z)$ pairs fall on a universal curve.  
We average the matched ITD, $Q(\nu,z^2,\mu^2)$, for cases where a given Ioffe time is obtained by different combinations of $(P_3,z)$. We denote such an average by $Q(\nu,\mu^2)$. The resulting fit is shown in the right panel of Fig.~\ref{fig:evolved_ITD}.

\begin{figure}[h!]
    \centering
    \includegraphics[scale=0.37]{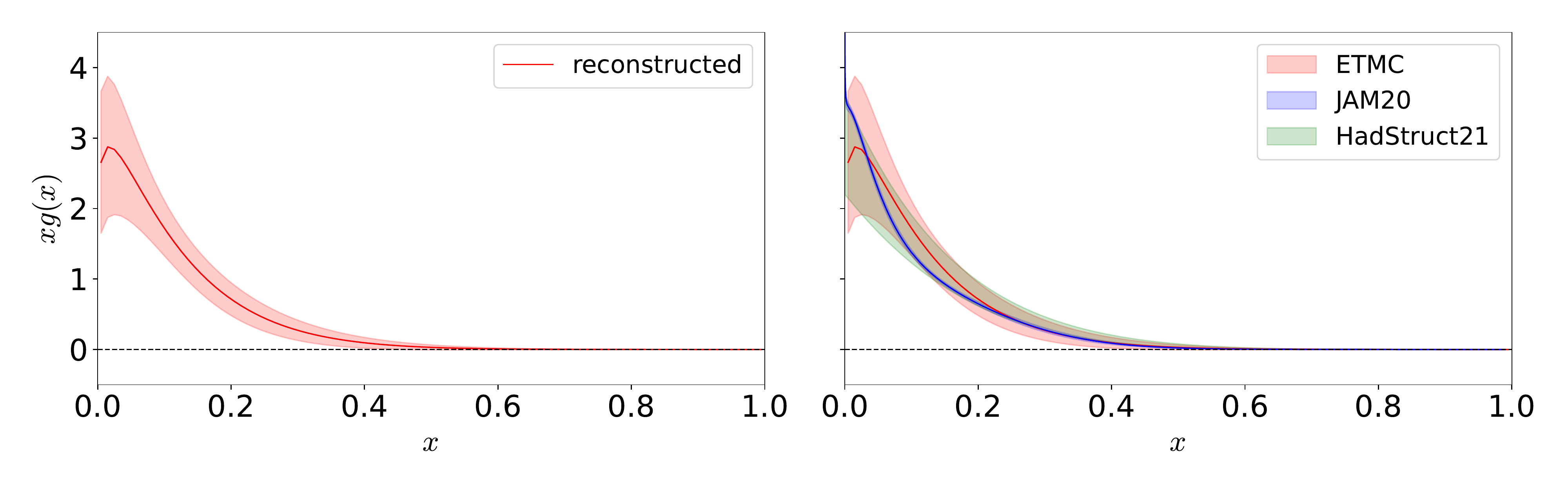}
        \vskip -0.5cm
    \caption{Left: The reconstructed gluon PDF. Right: A comparison of the PDF results from this work (red), JAM20 (blue), and HadStruc (green).}
    \label{fig:reconstructed_PDF}
\end{figure}

To extract the final gluon PDF, we use the fitting reconstruction, as explained in Sec.~\ref{sec:setup} and given in Eq.~\eqref{eq:ansatz}. The function is normalized using the gluon momentum fraction obtained with the same gluon and fermion action, but different lattice parameters, calculated in Ref.~\cite{Alexandrou:2020sml}. The reported value is $\langle x \rangle^{\rm \overline{MS}}_g(\mu=2 \rm{GeV})=0.427(92)$. The final result is shown in Fig.~\ref{fig:reconstructed_PDF}, where we find that the gluon PDF decays to zero faster than the quark contributions. In particular, $g(x)=0$ starting at about $x=0.4$. We also compare with recent results from the HadStruc Collaboration~\cite{HadStruc:2021wmh} and the global analysis JAM20~\cite{Moffat:2021dji}. The comparison is only qualitative, as the lattice results are obtained on a single ensemble. HadStruc uses an ensemble with similar volume and lattice spacing, $32^3 \times 64$, $a=0.094$ fm. Their source-sink time separation is also $9a$. The pion mass of the ensemble is $m_\pi=358$ MeV. In general, our results are consistent with the ones from HadStruc as well as JAM20. It is worth noting that the reconstruction performed by HadStruc includes values of Ioffe time up to $\nu_{max}=7.07$, while our reconstruction includes up to a maximum Ioffe time of $\nu_{max}=4.71$. Such a difference is attributed to two factors: (a) the use of the distillation method in the case of Ref.~\cite{HadStruc:2021wmh}; (b) the higher pion mass of their ensemble. Overall, the agreement between lattice results and global analysis is very promising.

\vspace*{-6mm}
\section{Summary}
\label{sec:summary}
\vspace*{-2mm}
In these proceedings, we presented results of the $x$-dependent unpolarized gluon PDF for the proton. The calculation was performed using an $N_f=2+1+1$ ensemble of clover-improved twisted mass fermions at a pion mass of about 260 MeV, a lattice spacing of 0.093 fm and volume $32^3\times64$. For the calculation, we employed the pseudo-distribution approach that significantly simplifies the renormalization procedure by forming ratios of matrix elements, leading to the reduced pseudo-ITD in terms of the Ioffe time, $\nu=z\cdot P$. In our calculation, we used nucleon momentum boosts up to 1.67 GeV and Wilson line length up to 0.56 fm, which suffices to extract a continuous dependence on $\nu$ and reconstruct the gluon PDF. We explored systematic effects such as excited-states effects using multiple source-sink time separations, the effect of stout smearing by comparing two values, as well as the dependence on the maximum value of $z$ entering the fits to obtain the ITD by testing $z_{\rm max}=6a,\,7a,\,8a$. For the evolution and conversion to the $\MSb$ scheme at a scale of 2 GeV, we used a one-loop formalism, ignoring the mixing with the quark singlet unpolarized PDF. Finally, we used the fitting reconstruction method to address the inverse problem and obtain the $x$-dependence of the gluon PDF. Our results were compared with other lattice data obtained using a different lattice formulation, methodology and setup~\cite{HadStruc:2021wmh} and we found very good agreement. Comparison with the global analysis of the JAM collaboration~\cite{Moffat:2021dji} is also very promising.

\vspace*{0.5cm}
\centerline{\textbf{Acknowledgements}}
\vspace*{0.15cm}

 J. D. and M.~C. acknowledge financial support from the U.S. Department of Energy, Office of Nuclear Physics, Early Career Award under Grant No.\ DE-SC0020405.
K.~C.\ is supported by the National Science Centre (Poland) grants SONATA BIS no.\ 2016/22/E/ST2/00013 and OPUS no.\ 2021/43/B/ST2/00497. 
Computations for this work were carried out in part on facilities of the USQCD Collaboration, which are funded by the Office of Science of the U.S. Department of Energy. It also includes calculations carried out on the HPC resources of Temple University, supported in part by the National Science Foundation through major research instrumentation grant number 1625061 and by the US Army Research Laboratory under contract number W911NF-16-2-0189. C.A. acknowledges financial support from the project EXCELLENCE/0421/0043 "3D-Nucleon," co-financed by the European Regional Development Fund and the Republic of Cyprus through the Research and Innovation Foundation, as well as from the EU project STIMULATE that received funding from the European Union's Horizon 2020 research and innovation program under grant agreement No. 76504.

\bibliographystyle{h-physrev}

\bibliography{references.bib}

\end{document}